17 Aug. 2024.    Update of the version of 27 Jan. with a mail from Michael Perryman in appendix SB10

# GaiaNIR: Note on processing and photometry

By Erik Høg, Niels Bohr Institute, Copenhagen University

**Abstract:** Some ideas for onboard processing and photometry with an astrometry satellite are presented, especially designed for GaiaNIR which may be launched about 2045 as a successor of Gaia. - Increased sensitivity, reduced image overlap, and simpler PSF calibration in GaiaNIR will result if the proposed initial processing of data from the detectors is implemented, because the across-scan smearing will become insignificant. - Filter photometry is required for high angular resolution as needed for astrometric and astrophysical reasons. Low-dispersion spectra are questioned because they fail at high star density. This will be a much greater problem with GaiaNIR than it is with Gaia because of the larger number of stars expected. It was the aim to collect in this note all arguments about GaiaNIR photometry which can be stated with words only, in correspondence with readers. The remaining work to be done for the definition of photometric equipment on the satellite requires further quantitative assessments and comparison of various options. Finally, 1) an advantage of filters is that the photometric observations can also be used for astrometry, 2) the XP spectra in Gaia will give very good astrophysical data for about 400 million single stars with G $<\sim$ 18.5 mag, but filters would have been better for all fainter and for all multiple stars, and 3) it is presently not clear which advantages for astrophysics low-dispersion spectra in the NIR might have over filters.

1. **Introduction**

   A Gaia successor was proposed by Høg (2013, 2014) and has since been developed to include near-infrared sensitivity as "GaiaNIR", presented by Hobbs et al. (2021, 2023). An industrial study of NIR detectors for the project is making progress, perhaps with sensitivity up to 2500 nm. This mission will be able to peer into the obscured regions of the Galaxy and measure up to 10 or 12 billion new objects in addition to two billion Gaia stars and reveal many new sciences in the process. ESA has ranked the development of this mission so high that it has good prospect of being realized and launched around 2045, see Høg (2021).

   Here follow sections 2-5 on processing and photometry and appendices SA and SB on the same two subjects.

2. **Processing onboard the satellite**

   The special detectors (Rixon et al. 2023) are fully read out at short intervals, dt, corresponding to the crossing time of one pixel. All pixels of each detector should be read to a *software* register R. Two more registers A1 and A2 are needed, about 20 elements longer along scan than R. R is added to each of these in the first part of them. The content of A1 and A2 is then shifted one pixel along scan and the first pixel in A1 and A2 is zeroed after the shift. The content is shifted across scan with the known speed in each of the two fields of view, FoVs, in the following manner.

   The extended parts of A1 and A2 beyond R with the accumulated data are not shifted across scan, but the left parts are shifted by one pixel when needed to compensate for the stellar motion, see figure 1. This has the effect that stars in each FoV appear as point sources in A1 or A2 while stars from the other field are smeared across scan. The problem with across-scan smearing known from Gaia will thus be minimized. The smearing is only one pixel as shown in A. Lennart says in section 4: "Reducing the across-scan smearing to an almost insignificant level will increase sensitivity, reduce image overlap, and simplify the PSF calibration."

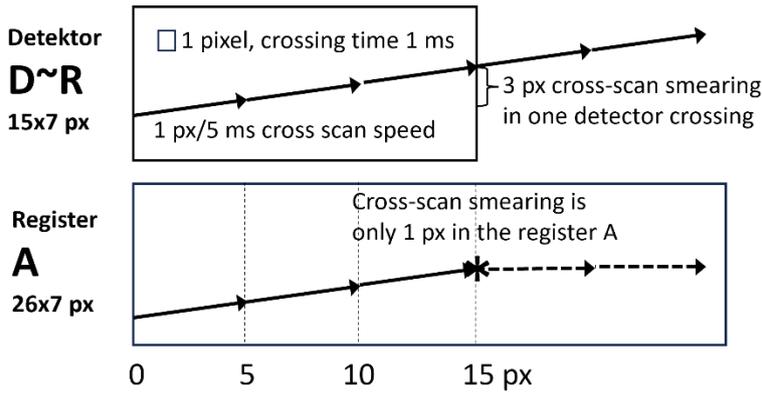

*Figure 1. The reduced across-scan smearing is shown. The detector D with the trace of a star, is here superposed on register R. The register A is shown with the accumulated image count, the count at the 15th pixel will be shifted along the dashed lines at right. The accumulated image followed the three traces indicated at left according to the specified shifts by one pixel across scan shown in figure 2 by Lennart. The assumed dimensions are unrealistic, but practical for the example, figure 3 shows more realistic sizes.*

The columns in A beyond the length of R contain the accumulated data, indicated by the dashed line, and shall be used for a windowing similarly to Gaia. Alternatively, the last column in the accumulating part of A could be treated like the readout register in a CCD and be read out for windowing as in Gaia. This alternative is probably preferred, and it is explained in detail in figure 3.

It is assumed that the first two detectors in the astrometric field, AF1 and AF2, where the stars enter are used for detection and verification of real stars, but we will not call them Sky Mappers, SMs, because they are not special, and they are used differently from Gaia. Only these detected objects are then windowed during the rest of the transit for transmission to ground, or for further processing onboard.

Lennart Lindegren sent the following diagram.

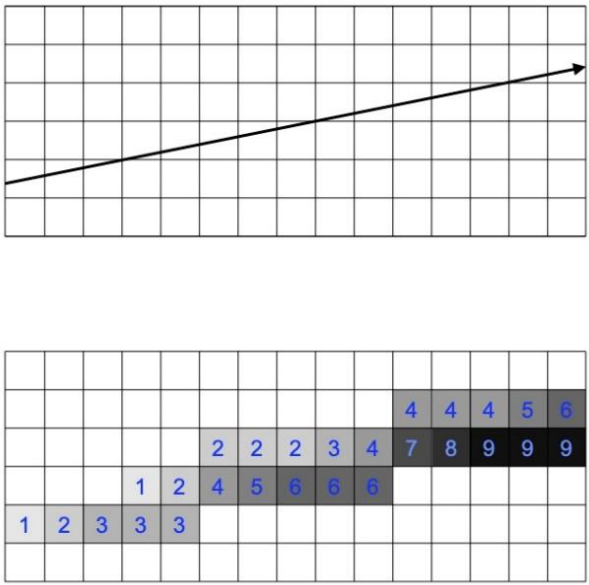

*Figure 2. Upper, the detector D with the stellar path. Lower, the register A with accumulated counts.*

Lennart's explanation to his diagram, figure 2:

I tried to make a slightly different visualization of how the counts are accumulated in register A, see attached figure. It shows in the upper diagram the path of the star image across the individual pixels of the detector (D), and in the lower diagram the corresponding counts in the A register. For simplicity the PSF is assumed to be much less than a pixel, and it generates one count per AL step. The AL shift in A is assumed to happen exactly when the star image crosses from one pixel to the next. This means that there is no smearing AL: at any time only one column of cells in A contains non-zero counts, and that column is shifted along with the image. However, for illustration purposes the diagram also shows the "old" counts (before they are shifted), rather as if the counts were copied instead of shifted. The AC shift happens after 5 and 10 AL steps. The grey shading increments by 10% per count.

For easy reference, let us label the pixels in D by their AL*AC coordinates, counted from the lower left corner, and the memory cells in A similarly by their AL*AC coordinates. At the end of step 5, only the 5th column of A contains non-zero counts, and they are 0,3,2,0,0,0 as shown in the diagram. At the transition to step 6, these counts are shifted one unit AL and one unit AC, so that they are now in the 6th column, which now reads 0,0,3,2,0,0. During step 6 the star image is in pixel (6,3) as shown in the top diagram, and one count is therefore added to cell (6,3) during that step. At the end of step 6, the 6th column reads 0,0,4,2,0,0. At the transition to step 7 they are shifted one unit AL and 0 units AC, so we have 0,0,4,2,0,0 in column 7 of A. The star is now in pixel (7,3), so at the end of step 7, column 7 reads 0,0,5,2,0,0, as shown in the bottom diagram. At the end of step 15, only column 15 of the A register contains counts, and reads 0,0,0,9,6,0.

Very bright stars will saturate the detector when they are integrated over the same time as faint stars, therefore shorter integration times must be implemented. In GaiaNIR this can be done alone with software without any hardware modifications of the detector as was required in Gaia. This is explained in figure 3 with more realistic sizes of the registers than in figure 1.

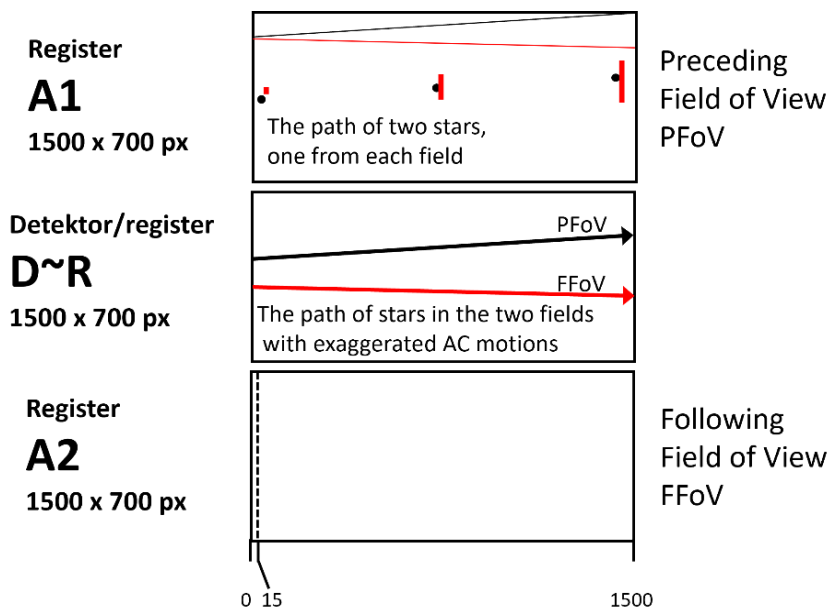

*Figure 3. Three software registers are proposed for onboard processing of the photon counts. All counts are read from the detector D after each basic time interval dt and stored in R. They are then added to the registers A1 and A2. The content of these is shifted in AL and AC as explained at the figures 1 and 2. An AC column thus contains all the accumulated counts from the preceding detector pixels. After the addition from R but before the shifting, the last AC column is read out and used for detection and windowing as in Gaia. The integration along the whole detector is appropriate for faint stars, but very bright stars will saturate the detector. Therefore, the counts at, e.g., the 15th AC column may be read out for detection and windowing, allowing 100 times brighter stars to be measured without saturation.*

In figure 3 the detector D shows the path of stars from the two fields of view. The register A1 for the P field shows the directions of motion by the lines at top. The accumulated counts are shown at three points of time for two stars in two colors, the star in black from the P field is a point thanks to the shifting, the other star from the F field is smeared in AC direction as shown in red.

The differential rate between the two fields is given by Lennart in section SA4 in a formula. With Gamma=106 deg we obtain a *maximum* of 275 mas/s, and with a spin rate of 60″/s the differential rate is 0.46%. With 1500 pixels AL and square pixels the differential motion in A of figure 3 becomes 6.9 px in a detector crossing. But the pixels will perhaps be 3 times wider in AC than AL as in Gaia, and Gaia has 4500 px AL. In that case the differential motion would be maximum 6.9 px in a crossing.

See more on processing in appendix SA.

## 3. Photometry

Filter photometry is required to obtain a high angular resolution, important especially for multiple stars and in dense fields. Whether 4 of 5 broad bands, must be decided from astrometric and astrophysical considerations. – The photometric system in the present GaiaNIR design is shown in Carrasco et al. (2023).

Low-dispersion spectra are questioned. Strong scientific justification is needed in view of the cost and complications, but the coming experience with the Gaia spectra may speak in favor of spectra as supplement to filters, not as substitute. But considerations in the following section SB2 show that a choice must be made for reasons of space in the focal plane.

The filter photometry is basic and indispensable because of the high angular resolution provided. It cannot be replaced by spectra because they suffer from too much confusion at multiple stars and in dense fields. This will be a much greater problem with GaiaNIR than it is with Gaia because of the larger number of stars expected.

Choice of the broad-band filters should be based on astrophysical arguments, but it should also satisfy the astrometric requirements for the vital correction of astrometric chromaticity. This correction should be obtained as directly as possible which is the case with filter photometry. Disentangling of overlapping spectra as in Gaia to derive the astrometric correction as explained in section 4, would however be acceptable with GaiaNIR but not for photometry.

Broad-band filters will give individual colors for multiple stars and in dense regions which is a great astrophysical advantage over spectra, and a sixth astrometric parameter in the AGIS solution will not be needed, or at least not as often as in Gaia.

The sixth parameter is a pseudo-color which may be used for some astrophysical purposes, and it can be calibrated to certain standard color indexes. But only one value is obtained for each star for the whole period chosen for the AGIS, i.e., one or several years. Filter photometry, however, will be obtained for each field transit and would be much more useful with the 4 or 5 magnitudes provided and because stars can be studied for both color and magnitude variability.

Experience with the BP/RP spectra from Gaia appears to be very good. No doubt on that, but this experience is of limited relevance for GaiaNIR because all Gaia sensitivity is at wavelengths less than 1.0 μm, while GaiaNIR will be sensitive from probably 0.8 to 2.5 μm. Proposals of spectra should be followed by assessment of the astrophysical value for specific science cases, and this should be compared with the performance obtained with the chosen filters for these cases.

## 4. Comments on photometry

Comment from Jos de Bruijne:

On the dispersion photometry that is used for Gaia, I am not 100% sure it is a bad idea also for GaiaNIR. The dispersion spectra were only released about 1 year ago and I have seen very ingenuous and powerful use of them. I think that we will be able to conclude in a few years' time, based on how scientists will use the data, whether the BP/RP spectra are powerful and useful or not.

Comment from Jos de Bruijne:

Spectrum overlap in dense regions endangers the (reliable) extraction of photometry for chromaticity calibration. This, however, is not a risk (or complexity) unique to GaiaNIR. This also applies to Gaia. Lennart and collaborators have therefore introduced a sixth astrometric parameter in AGIS (pseudo-color) that can be fitted in the astrometry, together with the 5 classical astrometric parameters, in those cases where the BP/RP spectra are not (yet) capable of providing reliable photometry. Whereas this is not a perfect solution, and the 5- and 6-parameter solutions have different error properties as a set, I consider it an acceptable solution which has only a (very) limited impact on the astrometric quality.

Comment from Lennart Lindegren:

I have no specific comments on the onboard processing in Section 2, except that I think it is an ingenious idea, if it can be implemented. Reducing the across-scan smearing to an almost insignificant level will increase sensitivity, reduce image overlap, and simplify the PSF calibration.

Concerning the photometry, I agree with your text on the sixth parameter. As Jos pointed out, the chromaticity problem for the astrometry can be overcome with the sixth parameter. This can be converted to a conventional color index, if needed, but the precision will be poor. Based on the experience from Gaia, the pseudo-color uncertainty is at least a factor 4 worse than the BP-RP photometry - which in turn may be less accurate than filter photometry would have been. Thus: only one color, no time resolution, and poor accuracy. It could do the job for the astrometry, but not for astrophysics.

Comment from Josep Manel Carrasco:

He agrees that the broad filter bands are required but emphasizes that in addition some narrow or intermediate width bands would be desirable. The choice of bands, the possible addition of spectroscopy, and possible use of Gaia BP/RP spectra should be carefully studied but this goes beyond the present note.

## 5. Conclusion on photometry

It was the aim to collect in this note all arguments about GaiaNIR photometry which can be stated with words only, it is thanks to open and extensive answers from the readers of seven successive versions of the note if this goal has been approached. The remaining work to be done for the definition of photometric equipment on the satellite requires further quantitative assessments and comparison of various options.

See more on photometry in appendix SB.

## 6. Acknowledgements

I am grateful for comments from Anthony Brown, Jos de Bruijne, Josep Manel Carrasco, Dafydd Wyn Evans, Claus Fabricius, David Hobbs, Óscar Jiménez Arranz, Carme Jordi, Arthur Kadela, Daisuke Kawata, Lennart Lindegren, Nicholas Rowell, and Michael Weiler. It was a great pleasure to attend the workshop in Lund (Hobbs 2023), to listen to and to talk with many colleagues. A presentation about across scan smearing, a talk with an expert about the detector, and the brainstorming about photometry have led to the present note, which was then developed in the correspondence with colleagues, see Høg (2023a).

# Appendix SA

Further comments on **photometry** were received in October 2023, see appendix SB.

Comments on **processing** up to 15 Sept.

SA1. Detection of stars
SA2. Windowing and transmission
SA3. Comment on 22 Aug from Nickolas Rowell
SA4. Comment on 27 Aug 2023 from Lennart Lindegren

### SA1. Detection of stars

It is assumed that the first two detectors in the astrometric field, AF1 and AF2, where the stars enter are used for detection and verification of real stars, but we will not call them Sky Mappers, SMs, because they are not special, and they are used differently from Gaia. Only these verified objects are windowed during the rest of the transit for data transmission to ground, or for further processing onboard. This may be done in the following manner, separately with data from the two registers A1 and A2 and separately for bright and fainter stars with data as shown in figure 3.

Detection is done in both AF1 and AF2 to discriminate between point sources and significantly elongated sources. The latter sources receive no further attention, they may be elongated because they are stars in the opposite FoV or they are due to errors of some kind. The point sources from AF1 having no counterpart in AF2 with a position within the uncertainty are neglected further on. All remaining point sources are treated as stars and are windowed in the further field transit, including the filter photometry. This data is transmitted to ground, unless an onboard processing shall take place which is considered below in section SA2.

This implies that some stars are sometimes windowed in both fields, namely at two periods of a satellite spin when they have nearly the same AC speed. A discrimination may be done on the ground. But the number of such cases could be much reduced if a detection is also done in the last astrometric field, AFL. The position of the point source from AFL will agree closely with the positions from AF1 and AF2 lest the source is in the opposite FoV and the differential AC rate between the two FoVs has moved the images apart, cf. section SA4 below.

Narrow double stars are of great astrophysical interest and deserve special attention. They might be missed in the detection process if the point-source requirement is very strict. The requirement should allow some elongation of the image especially in direction along the scan because treatment on the ground may well be able to disentangle the components. The detection should not require that the object is a *point source*, but that the detections in AF1 and AF2 at some level of confidence show that they have the same shape, same intensity and same position and therefore belong to a celestial object in the actual FoV.

If this new detection criterion is adopted, the onboard processing suggested in section SA2 cannot be maintained since all detections must be windowed. Or alternatively, a criterion might be defined to tell whether a windowing is useful or not.

The *detection process on Gaia* is recalled here, much guided by Jos de Bruijne:

There is an optical mask at beam-combiner level that ensures that SM1 only sees telescope 1 and that SM2 only sees telescope 2. So, there is on-board knowledge about the originating FoV for each object and this knowledge is actually used onboard for the AC propagation of the windows.

The two SM strips in Gaia are the same kind of CCDs as in AF, they are only sampled differently, namely 2x2 pixels for the whole CCD. These samples from SM1 and SM2 are used to detect objects. A detected object is accepted as an object in the sky if its presence is confirmed in AF1 *AND* its PSF shape in SM resembles that of a point source.

Gaia's PSF shape criterion is fairly generous such that narrow (unresolved or partially resolved) double stars would normally not be rejected,

 (see   https://ui.adsabs.harvard.edu/abs/2015A%26A...576A..74D/abstract ). Such cases obviously fit inside the same window.

The accepted objects are windowed for the rest of the field transit. The AC extent of the windows is constant and does not vary with magnitude (but the software on board in principle allows for different sizes). The AC extent is linked to the maximum AC motion and the windows are actually propagated in the AC direction during a CCD transit.

The cited paper by de Bruijne et al. (2015) says: "We find, as a result of the rectangular pixel size, that the minimum separation to resolve a close, equal-brightness double star is 0.23 arcsec in the along-scan and 0.70 arcsec in the across-scan direction, independent of the brightness of the primary."

This implies in case GaiaNIR implements the processing proposed above and if the pupil would have the same size as on Gaia (which it has not with the present design), that the minimum separation for such a double star would be the same along scan and perhaps 70% larger across scan as judged from figure 2, i.e., respectively 0.23 and 0.39 arcsec. But GaiaNIR will do a little worse compared to Gaia since the effective wavelength will undoubtedly be longer (redder) such that the PSF will be larger (wider).

The pupils on Gaia measure $1.45 \times 0.50$ m$^2$. For the CDF study of GaiaNIR 2 entrance pupils of 1.6x0.25m were used. These values are not fixed. The maximum AL size is 1.7m without involving expensive manufacturing procedures. It is still being considered if the AC size can be doubled to 0.5m.

### SA2. Windowing and transmission

All point sources in AF1 and AF2 shall be windowed. The windows from AF1 not accepted after verification in AF2 shall perhaps be rejected but let us see.

The following is *a possibility* to be considered later when we understand better how the detectors work and ideally, we should transmit as much window information as possible.

Since all accepted windows belong to point sources, it might be sufficient to derive some parameters to represent the astrometric and photometric information in the source. This should be done in one go with the windowing, and the number of parameters may depend on, e.g., the magnitude of the source. Only these parameters need transmission to ground. This would result in a saving on transmission but is it good enough for the science? What happens with the mapping around each source as was done with Gaia?

Ideally, the mathematical representation and the number of retained parameters in each case should be sufficient to recover on the ground almost the same information about the star and its surroundings that could be obtained if the full windows had been available. A compromise for the sake of mapping would be always to transmit full windows from some detectors, e.g., from AF1, AF2 and AFL.

Jos de Bruijne has sent this very welcome comment:

"If I understand this appendix correctly, the idea is to save telemetry by not downloading all samples of all windows but only some high-level parameters of fits to the window data that are made on board. This sounds problematic and a no-go to me. Gaia experience shows that the PSF model is color dependent, and this calibration is continuously improved in Gaia from one data release to the next. I believe that Gaia experience has demonstrated beyond doubt that the raw sample data, acquired on board, need to be transmitted to ground (with lossless compression). I deem any scientific processing on board a great risk for the astrometric performance of the mission. As now demonstrated by Euclid, we should use K-band downlink of the science telemetry which is a factor 7.5 more powerful compared to Gaia's X-band."

### SA3. Comment on 22 Aug. from Nickolas Rowell
### about the proposed processing

Dear Erik,

Thanks for sharing this note. I gave the talk on AC smearing in Lund, so it's good to see that this issue is being considered seriously and seems to have an elegant solution thanks to the improvements in detector technology and onboard processing capacity. I also developed the models of drift-scan effects (across and along scan) in Gaia's LSF & PSF, and they're currently being used for DR4 processing. The details are still only written up in internal technical notes, but I plan to write a paper later this year describing the models.

I have a few comments and observations:

 - I suppose this procedure will limit the AC smearing to 1 pixel, rather than eliminate it entirely; there will still be a small AC-rate dependent modulation in the PSF, but most of the effects that impact Gaia observations will be greatly reduced (including the interaction between AC rate and detector spatial response variations).

- Using detectors that are much shorter in the AL direction will reduce the AC smearing even in the absence of any special processing.

- The spacecraft needs to know the instantaneous AC drift rate onboard in real time with sufficient accuracy; there may be a small fraction of observations affected by attitude disturbances for which the proposed processing fails - or maybe the AOCS will be able to track this accurately enough.

- There will still be along-scan smearing due to differences between the TDI rate and the along-scan rate; this is less important because it doesn't correlate with the parallax factor in the same way as the AC smearing and can be modelled in the LSF/PSF as we currently do for Gaia. Having shorter AL detectors will reduce the impact of this anyway.

Regards,  Nick Rowell

**SA4. Comment from Lennart Lindegren on 27 Aug.**
**about the smearing of images from the opposite field**

The AC shifting of the A registers will reduce the AC smearing to < 1 pixel for one of the fields (e.g. the P field in A1, and the F field in A2). Images from the opposite field (F in A1, and P in A2) will be smeared by the differential AC rate, which is 2*S*sin(Gamma/2)*cos(Omega), if S is the amplitude of the AC rate in one field, Gamma the basic angle, and Omega the heliotropic spin phase. The differential AC rate thus varies sinusoidally over a spin with an amplitude of 2*S*sin(Gamma/2). For Gaia, and in fact any reasonable scanning law, S = 170 mas/s as required for good sky coverage. Thus the overlap problem caused by images in the opposite field essentially increases with the basic angle as sin(Gamma/2). This could be an argument for not making the basic angle unnecessarily large.

# Appendix SB

**Comments on the photometry** where I am grateful for

comments from Anthony Brown, Dafydd Evans, Claus Fabricius, David Hobbs, Carme Jordi, and Michael Weiler:

SB1. Comment on 1 Oct. by Anthony Brown

SB2. Comment by Erik Høg

SB3. Comment on 13 Oct. by Anthony Brown

SB4. Comment by Erik Høg

SB5. On Gaia photometry and the decision in 2006

SB6. Comment on 6 Nov. by Michael Weiler

SB7. Comment on 10 Nov. by Dafydd Evans

SB8. Final remarks on the design of photometry

SB9. Comments on 15 Nov. by David and Erik

SB10. Comments on 17 Aug. 2024 from Michael Perryman and Erik

**SB1. Comment on 1 Oct. by Anthony Brown**

Here are some comments from my side on the question of filter photometry for GaiaNIR vs low resolution spectra as used for Gaia. I copied David as these notes may also be useful for him.

Best regards,   Anthony

I am convinced that since the release of Gaia DR3 it is clear that the low resolution BP/RP spectra are a crucial element in delivering the full power of the science that can be done with a Gaia-type survey. In particular in disentangling the formation history of the Milky the determination of precise astrophysical parameters as well as radial velocities is essential. Distance and kinematic information are simply not enough. A recent work demonstrating this is by Horta et al. https://arxiv.org/abs/2307.15741 . In particular their figures 11 and 12 show the great difficulty of unraveling the Milky Way's history without precise astrophysical information.

Various works have now shown that the Gaia XP spectra, especially in combination with infrared photometric surveys contain precise and robust information on metallicity and alpha-element enhancements, two key parameters for studying the Milky Way. Examples include Andrea et al. https://ui.adsabs.harvard.edu/abs/2023ApJS..267....8A/abstract , Jiadong Li et al. https://arxiv.org/abs/2309.14294 , Martin et al. https://arxiv.org/abs/2308.01344 , and Xiangyu Zhang et al. https://academic.oup.com/mnras/article/524/2/1855/7209172 . Guiglion et al provide an example of combining Gaia parallaxes, photometry, XP and RVS spectra to derive astrophysical parameters https://arxiv.org/abs/2306.05086 .

As one of the highlights, the information in XP spectra has allowed tracing the first fragments of the Milky Way in formation, some 12.5 billion years ago, see Rix et al. https://ui.adsabs.harvard.edu/abs/2022ApJ...941...45R/abstract .

The additional advantage of low resolution spectra is that one can synthesise specific photometric bands as shown in Gaia Collaboration, Montegriffo et al. https://ui.adsabs.harvard.edu/abs/2023A%26A...674A..33G/abstract and used in the paper by Martin et al cited above.

The above works also pave the way for much improved methods (over those used by DPAC) for the extraction of astrophysical information from a Gaia-type survey. This will of course be used in a future analysis of the processed GaiaNIR data.

While I agree that for the purposes of achieving the best resolution for a survey of stellar colours (for astrometric data processing) filter band photometry is preferred, I think we should very seriously consider a combination of filter bands and low-resolution spectra. A well-chosen set of filters can provide the necessary colour information for the astrometry and can enhance the power of the low resolution spectra (see references above). For cost reasons it may well be necessary to make a choice, but in that case again I would argue we should seriously look into the low-resolution spectra option.

Looking at the broader context, by the time GaiaNIR would fly, the Euclid and Nancy Grace Roman space near-IR surveys will have completed. The mirror sizes of these missions are likely to remain larger than the GaiaNIR mirrors and these surveys are likely to remain deeper than GaiaNIR. While not all-sky (although large fractions of the sky will be covered), the data from Euclid and Roman can certainly be used as part of the GaiaNIR processing and may thus help in, for example, the PSF calibrations. The Euclid mission in fact makes use of the XP spectra to aid the PSF calibration for their survey in the visible.

## SB2. Comment on 9 Oct. by Erik Høg

The thoroughly documented arguments by Anthony show that low-dispersion spectra must be very seriously considered for GaiaNIR. I still think, however, as I wrote in the section 3 that the remaining work to be done for the definition of photometric equipment on the satellite requires further quantitative assessments and comparison of various options.

In section 3 I wrote: "Filter photometry is required to obtain a high angular resolution, important especially for multiple stars and in dense fields. Whether 4 of 5 broad bands, must be decided from astrometric and astrophysical considerations. Low-dispersion spectra are questioned. Strong scientific justification is needed in view of the cost and complications, but the coming experience with the Gaia spectra may speak in favor of spectra as supplement to filters, not as substitute." But the following considerations show that a choice must be made for reasons of space in the focal plane. Filter photometry is the only way to get astrophysical information of the stellar components at high star density.

A colleague from Gaia recently wrote to me that it was a mistake to use spectra in Gaia in his comment to my review paper (Høg 2023b). I answered, that in a way I agree but that we had too little time to think when that decision was made, and I decided to dig deeper with the present result of the very thorough comment from Anthony who has taken the time needed despite the busy time he must have as leader of the DPAC consortium.

The discussion of Gaia photometry at the end of section 17 in my review paper is copied to the following section SB5:

The space needed in the focal plane is an issue. The Gaia focal plane may be taken as a concrete example for this discussion, see figure 17 in the review paper. The space presently occupied by the XP spectra would be sufficient to accommodate 3 strips for filters if the filters can be evaporated on to the detectors. This is feasible nowadays, but glass filters would have been used when Gaia was built and extra space between the strips would have been needed. The 5-filter photometry proposed in table 3 of Jordi et al. (2006) is placed on 4 strips since two colors share a strip. With 4 strips for photometry, one strip would be taken from astrometry which seems acceptable.

If filter photometry and XP spectra were implemented together, the 4 strips would be taken from astrometry, and even more than 4 strips in the case of glass filters. Therefore, a choice seems unavoidable. But the present discussion does, in my view, not imply that we should have adopted glass filters in 2006, neither together with spectra nor instead of spectra - if time had permitted.

We really want to discuss the choice between spectra and filters for *GaiaNIR*. But since reliable data for the special NIR detectors about noise and sensitivity is not yet available it may be considered for the present to study the choice for a *Gaia twin* as a concrete example where all performance data is available.

It is not evident as far as I see from section SB1, whether spectra are better than filters Clearly, spectra are very good but are they better than filters? This question could be studied for one of the papers mentioned by Anthony comparing the outcome if the Gaia twin had filters on the detectors and if it had XP spectra.

Finally, 1) an advantage of filters is that the photometric observations can also be used for astrometry, 2) the XP spectra in Gaia have given very good results, but it is not known if 4- or 5-color filters would have been better, and 3) it is presently not clear which advantages for astrophysics low-dispersion spectra in the NIR might have over filters.

## SB3. Comment on 13 Oct. by Anthony Brown

I have no further comments on the paper.

On the note about photometry, I want to insist that we have to study including low dispersion spectra. I agree that for resolution reasons it is optimal to have filters, but perhaps the combination of filters and spectra can allow for economizing on both aspects. What I have in mind is that the spectra provide information on the stellar SEDs over the full wavelength range and perhaps we can live with a smaller set of cleverly chosen filters which provide the astrometric colour information as well as the information needed to better disentangle overlapping spectra.

I don't think there is any doubt about the astrophysical information in the spectra. We know from surveys like APOGEE and the ongoing SDSS-V that the infrared spectral range contains very rich information on metallicity and alpha-element enhancements. I am convinced this information can also be extracted from lower resolution spectra, as already proven for the optical.

best regards,                    Anthony

## SB4. Comment by Erik Høg

My conclusion as before: Inclusion of low dispersion spectra and of filters shall both be studied. Whether both should finally be included is a matter to be decided in the context of the whole satellite design. It is important to know which option shall be chosen if such a choice is requested at short notice as happened in 2006 during the final development of Gaia, see the following section SB5 with an extract from Høg (2023). To date, we know that the Gaia spectra have performed very well for astrophysical studies of single stars but let us now compare with filters.

The photometric accuracy with filters should be better since the background and read noise will be smaller than for the corresponding spectral band in a spectrum because the signal in the filter band is collected from fewer pixels and a smaller area of the sky than in the spectrum. This will be very significant for faint stars. Studies where only the BP/RP photometry is available would be better off with pure filter photometry in 4-5 bands. The low-dispersion spectra can only be an advantage if the higher spectral resolution can be utilized. It would be interesting with this in mind to look at the 8 references given by Anthony in SB1. I have tried but it is not easy to find those using the XP spectra.

The Gaia DR3 (see section 5.3.6 in the documentation) contains mean spectra for 219 million sources, most of them with G < 17.6 mag, from data collected in 34 months. The 10-year mission may have 400 million mean spectra for sources brighter than G <~ 18.5 mag, and these spectra will serve astrophysics better than filters would have done. The remaining fainter 1.4 billion stars will get BP, RP and G magnitudes, it would have been better for them and for all multiple stars if Gaia had 4-5 band filters. This is perhaps a fair comparison of spectra and filters for Gaia, and it quantifies the dilemma of a choice between spectra and filters.

Anthony commented: "This is fine."

He added soon after: "Yes, okay. Just to be clear, my comment does not mean I fully agree with what you state. Certainly, more investigation is needed, and I would still argue for the low-resolution spectra option (if one had to choose)."

I believe that the problem is not to make the spectra, but a question is whether the faintest of these spectra contain more information than the photometry.

Michael Weiler in Barcelona sent his opinions copied here in SB6.

The spectra are made by Dafydd Evans and his group in Cambridge. Dafydd's comment is placed in SB7.

Final remarks follow in SB8.

Comments on great news by David and Erik in SB9.

Claus Fabricius in Barcelona has responsibility for the validation of the successive versions of the Gaia Data Releases, and he helped me in various ways with the present note on photometry.

Data for the special NIR detectors about noise and sensitivity is not yet available, and it may take several years before the detector development is so far that reliable data can be provided. It is therefore suggested for the present to study the choice for a *Gaia twin* as a concrete example where all performance data is available. Such a study would be a good preparation to the study of this matter in GaiaNIR.

## SB5. On Gaia photometry and the decision in 2006

An extract from section 17 in Høg (2023) follows. The main point here is to show how the decision in 2006 of replacing filter photometry with the low-dispersion spectra in Gaia had to be taken on very short notice, without time to study alternatives. To date, we know that spectra have performed very well, but *we do not know if filters would have been better*, even at single stars.

This should be a warning to be ready to make a similar choice for GaiaNIR: For the case a choice must be made at some time, the solid arguments must be ready.

"…Another of my major activities in optimizing the Gaia payload was on the photometry. From the earliest ideas for Gaia, it was very clear that measuring a billion or more objects was made far more powerful if we could also record their colors or spectra. Photometry was important, in fact for two different reasons. The instrument gives small but significant chromatic errors in the astrometry which can be corrected if a color index of the star is known, and we need to know that for every transit for the sake of variable stars with changing colors. The other reason is astrophysical: the color gives its position in the Hertzsprung-Russell diagram, and (if the filters are carefully chosen) also the spectral type, metallicity, and reddening. Many other photometric surveys are going on, but we needed a reliable source giving prompt results: So, it had to be by Gaia itself.

There were several key people helping to think through the objectives, and the possible solutions. My colleague in Copenhagen, Jens Knude, was my constant support and he brought his lifelong dedication to photometry, especially with the Strömgren uvby-system with intermediate width bands suited for early type stars. People in Lithuania were also keen to support, in fact Professor Vytas Straizys wrote to me already in 1994. He was director of the Vilnius Observatory and had developed the "Stromvil" system suited for classification of all spectral types, also in the presence of interstellar reddening. So, the experts joined our efforts very early. For photometry and spectroscopy, the first idea was to add a third telescope with a collecting aperture of 0.75 x 0.70 m² in addition to the two large telescopes for astrometry. This was the design by Matra Marconi Space at the end of the industrial study in mid-1998.

Various possibilities were considered, ranging from multiple filters to low-resolution spectroscopy. When the cornerstone study started, a Gaia Science Team was formed in 2001, under Michael Perryman's leadership, with various working groups devoted to studying specific aspects. One of these was the photometry working group with Carme Jordi from Barcelona and me as co-leaders. We had many special meetings on photometry and wrote many reports in those years. In the first design, photometry in four broad bands was obtained in the astrometric telescopes and in seven medium-width spectral bands in the smaller telescope. The next design, which was described in detail in The Concept and Study Report, ESA (2000b), increased this to eleven intermediate bands. In 2006 our work ended with a paper of 25 pages in the journal Monthly Notices of the RAS with 35 authors. The number of bands had increased to a total of 19 optimized for astrophysics.

But that wasn't the end of this long and important story. NO! All this changed dramatically. In fact, it changed while we were finalizing the MNRAS paper, but only a few of us knew that. The change had to be kept confidential because it was taking place during the competitive bid for the industrial contract for Gaia. One of the competitors, Matra Marconi Space (who eventually became the winning contractor), was proposing to simplify the design for reasons of mass and cost. They dropped the dedicated photometric telescope and reduced the number of focal planes from four to only one.

They asked Michael whether it would be acceptable to replace photometric filters with two low-dispersion prisms. They reluctantly accepted that he also engaged me, and I also wanted to have Carme Jordi, Anthony Brown from Leiden, and Lennart Lindegren involved. It was a complicated problem, but eventually a very careful optimization of the prism properties was done utilizing the MNRAS paper, resulting in the very satisfactory Gaia photometry that we have today.

I believe, however, that our decision would have been different if we had known much earlier that the dedicated photometric telescope had to be dropped. Given more time to think, we would probably have chosen multi-color (4 or 5) filter photometry for Gaia because the resulting high spatial resolution would give better photometry at multiple stars and in regions with high star density and the photometric accuracy would be better because of the lower background than with spectra. But that was not an option in 2006 at this critical point in the development. A 4-color broad-band photometry was in fact included in the design on figure 3.8 of the astrometric field in ESA (2000b) and this was improved to five bands in the final report on Gaia photometry (Jordi et al. 2006). – For a future mission as discussed in section 20, a combination of filter photometry and spectra is perhaps optimal in view of the very good experience with Gaia photometry, but this can hardly be accepted because the extra space needed in the focal plane if both filters and spectra are included would necessarily be taken from astrometry. Furthermore, 1) an advantage of filters is that the photometric observations can also be used for astrometry and 2) it is presently not clear which advantages for astrophysics low-dispersion spectra in the NIR might have over filters.

The Gaia mission was developed, the satellite was launched in December 2013 and this wonderful instrument is shown in the figures 17, 18 and 19…"

## SB6. Comment on 6 Nov. by Michael Weiler

Michael Weiler is involved in the development of the calibration models and the validation for XP spectra in Barcelona. He sent his opinions copied here with 1) general considerations on spectra versus filters, 2) specific ideas on the design of low-dispersion spectroscopy taking calibration into account, 3) he notes that mean spectra for all sources are being produced already to be released later, and 4) on the faintest spectra.

1) Concerning the choice between low resolution spectra in Gaia XP style, or filter photometry, I think the situation is complex. Spectroscopy even at very low resolution allows for the study of specific features in the spectra better than filter photometry. This is something that in my opinion the work with XP spectra until now already has demonstrated. But low-resolution spectroscopy also has some disadvantages. Because of their larger size in the focal plane, they are much more affected by crowding in dense regions of the sky than filter photometry is. Also, the signal-to-noise ratio for faint objects might become lower for faint sources, when the signal is spread over more pixels in spectroscopy, than it is in filter photometry. So, the choice between the two options is not obvious to me. When being interested in absorption and emission lines in relatively bright objects I personally would go for low resolution spectra rather than filter photometry. For classifying faint objects in dense regions, I would prefer photometry in 4 to 5 bands, though. Probably one would have to make a clear, prioritized list of objectives, and then investigate in detail, probably with simulations, if low resolution spectra or filter photometry is the overall the better choice...

2) Let me also add that I don't think that the adoption of low-resolution spectra for Gaia has been a mistake. But I think that the design of the spectra has been sub-optimal. When designing the instruments, not much thought about the calibration of the spectra seems to have entered the process. When designing a new low-resolution instrument, I would introduce some modifications with respect to the Gaia XP instruments because of the experiences from calibration. In particular I would change the design in such a way that the changes in the overall response (i.e., the combined effect of detector quantum efficiency, optical transmission, and filter transmission) are much smaller on scales of the width of the LSF. Doing so would dramatically simplify the calibration of the spectra, as it would allow for the separation of the response function from the LSF. So, no wavelength cut-on and off-filters anymore.

3) As a final remark, the text in the document says that there will be ~400M XP spectra for sources brighter than 18.5. But mean spectra for all sources are being produced already, and as far as I know will also be released some day.

4) To the question whether the faintest of these spectra contain more information than the photometry. For the faintest sources the low-resolution spectra will be very noisy, and the faintest sources will be the most frequent ones. One might think of binning to increase the signal-to-noise ratio. But the read-out noise might still be larger for binned spectra as compared to filter photometry. And binning only over some wavelength intervals, to maintain some colour information, can result in different effective passbands for sources of different colour. This is related to the problem of calibration if the total response changes rapidly as compared to the LSF mentioned before.

## SB7. Comment on 10 Nov. by Dafydd Evans

I asked Dafydd:

"Could you perhaps add something about the astrophysical value of the faint spectra about G=17-18, compared to 4-5 color photometry if that had been the choice for Gaia? I believe that filters would be definitely better for multiple stars because the components would be separated. But what about single stars at these magnitudes? Would the astrophysics with filters also be better than spectra because of the lower noise?"

Dafydd does not mention my question in his answer where he writes: 1) About Jos's comment in section SA2 with which I also agree. 2) About the choice between low-dispersion spectra and filters. 3) On how to proceed with the design. 4) He would prefer to include both in GaiaNIR.

1) I agree with Jos's comment regarding doing some high-level fits on-board the satellite. This would not be a good idea since you can't reprocess the data. CU5/DU10 have improved the LSF/PSF determination considerably over the Gaia mission.

2) Regarding the choice between low-dispersion spectra and filters, I think that the answer is complex (as most others have said). If you had asked me earlier in the year, I would definitely have argued for the filters. The calibration is much, much simpler and they will not have as many crowding problems. GaiaNIR's big selling point is that it is going into these crowded regions in the Galactic plane. However, after attending the Gaia XPloration, my eyes were opened to the amount of information that was being extracted from the Gaia XP spectra. Users were very excited with this data and amazingly ingenious (usually in a machine-learning way) as to how they were extracting the wealth of data from the spectra. The big advantage of the spectra is the flexibility that they bring. I think that it would still be necessary to have a small filter set to get some colour information in the crowded regions.

3) The only way to really work out the correct solution is to do a large-scale simulation combined with astrophysical extraction using both types of data. You would then have to make a judgement on the results of the performance as a function of magnitude and crowding level. That would not be easy.

4)  My instinct is that you need a combination of low-dispersion spectra and filters.

## SB8. Final remarks on the design of photometry

The question I asked Dafydd in SB7 should be answered.

It appears that there is a preference for including both low-dispersion spectra and filters.

I agree that this is a good starting point for the design. But I think it is prudent to be able to make a choice even at some late point, forced, e.g., by arguments of mission cost. This is not only prudent, but even a must for an optimal design.

The astrophysics of dense regions and of double and multiple stars will greatly benefit by color information which can only be obtained with filters. The performance of the spectra for these cases should be studied for comparison. How do these cases fare in Gaia?

Dafydd received these remarks with an immediate reply: *"That would be fine. We are in an early enough stage of the mission to try and get funding for someone to do the performance investigation."*

## SB9. Comments on 15 Nov. by David and Erik

David wrote on 15 Nov.:

For the photometry it was clear from the meeting in Lund and in your notes that both filter and spectral photometry are desirable for different reasons. More detailed studies of a new design for GaiaNIR are ongoing but I do not see any major problems in including both especially as GaiaNIR is now an L-class mission and a greater science return is required.

I replied:

This is news for me, very good news, that GaiaNIR is definitely L-class. As I wrote, the inclusion of filters and low-dispersion spectra is a good starting point for the design. And I still think it is prudent to be able to make a choice even at some late point, forced, e.g., by arguments of mission cost. So, the astrometric and photometric performance of options only with filters and only with spectra should be calculated and be compared with the filters + spectra option."

It is a pity we do not have such a comparative study of filters and spectra for Gaia. In section SB4, I proposed to make it soon. To date, we know that spectra have performed very well, but *we do not know if filters would have been better*, also for astrophysics of single stars.

## SB10. Comments on 17 Aug. 2024 from Michael Perryman and Erik

Michael wrote on 12 May:

Dear Erik,

There is a recent paper that I thought you would find very interesting: you may be aware of it already. The paper is by Montegriffo et al. 2023, and titled: "The Galaxy in your preferred colours: synthetic photometry from Gaia low-resolution spectra"

https://ui.adsabs.harvard.edu/abs/2023A%26A...674A..33G/abstract

They use the mean BP/RP low-resolution spectra published as part of DR3, and through appropriate calibration of any specified photometric passbands, they can replicate the photometric data acquired in those passbands.

Amongst narrow-band systems, they detail results for the Stromgren system and also for the proposed broad- and medium-band filters for Gaia (Section 4.4). They showed that they could, for example, clearly separate giant and main sequence stars in the log g versus Teff plane (their Figure 22), and could pick out both the white dwarf sequences, and the Jao gap in the various colour-magnitude diagrams. So, the C1 filter system was clearly very well designed. Carme was a co-author on the paper, so she was probably responsible for this part.

In their earlier paper detailing the external calibration of the BP/P spectra https://ui.adsabs.harvard.edu/abs/2023A%26A...674A...3M/abstract they also showed that the Hipparcos/Tycho photometry are all reproduced by the BP/RP synthetic photometry to better than 2.5 milli-mag.

I thought these results were all very interesting, and thought that you would enjoy reading about it too.

The low-resolution spectra have clearly proven to be spectacularly successful, and I will go into more details in my future essays 187 and 189.

kind regards    Michael

Erik wrote: Hi Michael,

Thank you for the very interesting information, I have read the papers and the two new essays. The results are very impressive. I have understood the low-dispersion spectra have been very successful, but I still think that a careful study is required to decide for a future mission whether filters should be preferred. However, the spectra for *bright stars* may have advantages over filters, see also SB6 by Michael Weiler.

I have discussed the photometric system for GaiaNIR with several people in the present report. I would like to hear your opinion.

Michael answered: "Excuse me from not getting into the issues of the GaiaNIR photometry. I have not been following the science case, and could not make a useful contribution without investing more time than I have at present."